\documentclass[conference]{IEEEtran}
\IEEEoverridecommandlockouts
\usepackage{cite}
\pdfoutput=1

\usepackage{float}
\usepackage{booktabs}
\usepackage{multirow}
\usepackage{multicol}

\usepackage{amsmath,amssymb,amsfonts}
\usepackage{algorithmic}
\usepackage{graphicx}
\usepackage{textcomp}
\usepackage{xcolor}

\usepackage{algorithmic}
\usepackage{array}
\usepackage[caption=false,font=normalsize,
   labelfont=sf,textfont=sf]{subfig}
\usepackage{stfloats}
\usepackage{url}
\usepackage{verbatim}
\usepackage{balance}

\usepackage{colortbl}  
\def\BibTeX{{\rm B\kern-.05em{\sc i\kern-.025em b}\kern-.08em
    T\kern-.1667em\lower.7ex\hbox{E}\kern-.125emX}}

\begin{document}

\title{Enhancing 1-Second 3D SELD Performance with Filter Bank Analysis and SCConv Integration in CST-Former\\
\thanks{This work was supported by the Department of Computer Science and Engineering at Southern University of Science and Technology.}
}
\author{
  Zhehui Zhang \\
  Department of Computer Science and Engineering \\
  Southern University of Science and Technology \\
  Shenzhen, China \\
  \texttt{zzh953929932@gmail.com}
}

\maketitle

\begin{abstract}
Recent SELD research has predominantly focused on long-time segment scenarios (typically 5 to 10 seconds, occasionally 2 seconds), improving benchmark performance but lacking the temporal granularity needed for real-world applications.
To bridge this gap, this paper investigates SELD with distance estimation (3D SELD) systems under short time segments, specifically targeting a 1-second window, establishing a new baseline for practical 3D SELD applicability.

We further explore the impact of different filter banks---Bark, Mel, and Gammatone for audio feature extraction, and experimental results demonstrate that the Gammatone filter achieves the highest overall accuracy in this context.
Finally, we propose replacing the convolutional modules within the CST-Former, a competitive SELD architecture, with the SCConv module. 
This adjustment yields measurable F-score gains in short-segment scenarios, 
underscoring SCConv's potential to improve spatial and channel feature representation. 
The experimental results highlight our approach as a significant step towards the real-world deployment of 3D SELD systems under low-latency constraints.

\end{abstract}

\medskip


\section{Introduction}
\begin{figure}[t]  
    \centering  
    \includegraphics[width=0.4\textwidth]{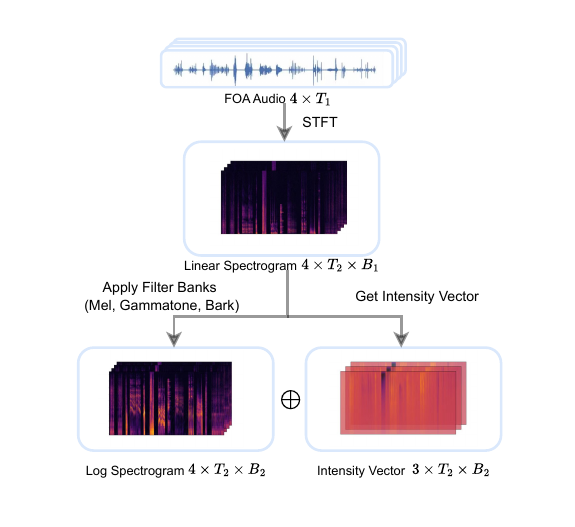}  
    \caption{Flowchart of the data preprocessing pipeline. The arrows denote non-parametric operations, while the blue boxes represent different data segments. The symbol $\oplus$ indicates the concatenation of extracted features, integrating them into the final input representation. In this figure, $T_i$ represents the time-dimension frames, and $B_i$ refers to the frequency-dimension count.}
    \label{fig:preprocessing}  %
\end{figure}
Computational Auditory Scene Analysis (CASA) focuses on developing machine listening systems replicating human auditory perception in complex acoustic scenarios \cite{CASA}. As a pivotal subset of this field, Sound Event Localization and Detection (SELD) seeks to enhance the automated identification, localization, and tracking of multiple concurrent sound events.

\par SELD systems are utilized across diverse applications, ranging from robotic navigation to smart room acoustics optimization \cite{smartroom}. 
SELD systems are essential for real-time spatial audio applications, such as advanced teleconferencing and context-aware wearable devices \cite{wearable}, which demand precise localization and low-latency responses. 
This requirement necessitates algorithms that ensure precise spatial-temporal resolution under low-latency constraints.
Existing SELD research has predominantly focused on enhancing localization accuracy using long time segments (5-10 seconds, occasionally 2 seconds), which limits applicability to dynamic real-world scenarios requiring finer temporal resolution. 
While multi-ACCDOA \cite{multiaccdoa} explores a 1.27-second segment length, it falls short of explicitly addressing the need for higher temporal granularity in real-time applications. Previous real-time systems focused on separate sound event detection (SED) \cite{realsed0}\cite{realsed1}\cite{realsed2} and direction-of-arrival (DOA)\cite{realtimedoa0}\cite{realtimedoa1}\cite{realtimedoa2} tasks. 
SELD-Edge \cite{seldedge} tested 1-second audio segments for edge device applications but excluded distance estimation. 
To bridge this gap, our study evaluates 3D SELD performance using a 1-second window, establishing a new benchmark for short-segment analysis.
\begin{figure*}[t]  
    \centering  
    \includegraphics[width=0.9\textwidth]{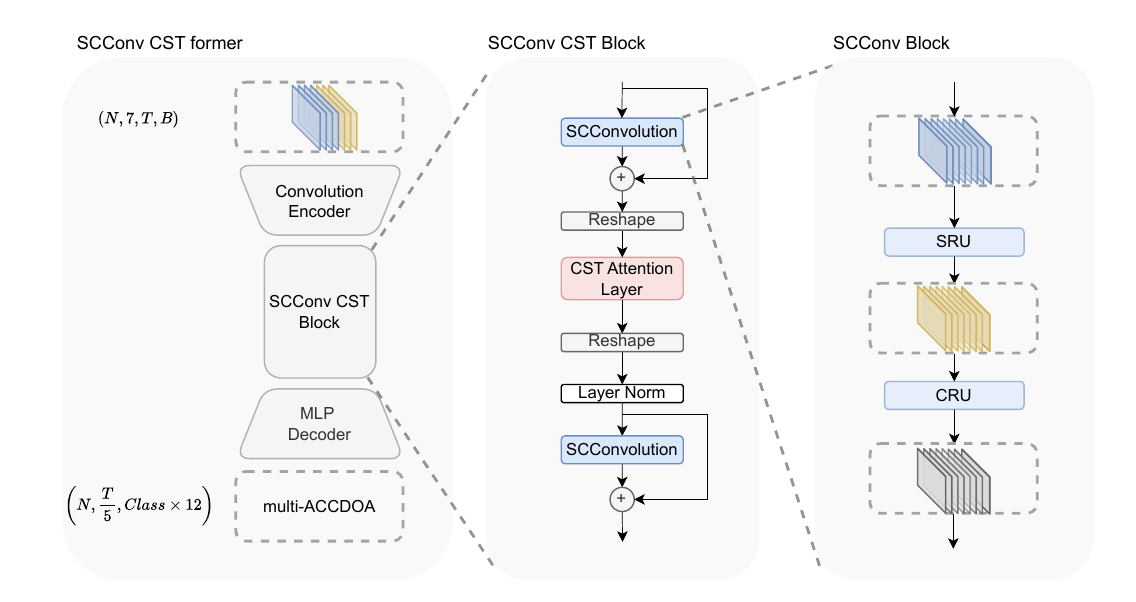}  
    \caption{Overview of the SCConv CST-former architecture. The left panel presents the full model, while the middle and right panels show the detailed designs of the SCConv CST and SCConv blocks. The SCConv block, replacing the original Local Perception Unit and Inverted Residual FNN in the CST block, improves local feature extraction and fusion. Here, $N$ represents the batch size, $T$ is the time-dimension frame count, $B$ refers to the frequency-dimension frame count, and $Class$ denotes the total number of sound event classes.} 
    \label{fig:model}  
\end{figure*}
\par SELD research has made significant advancements in network architecture design for both baseline and competitive models. 
The initial SELDnet \cite{2018seldnet} used convolutional layers for feature extraction and gated recurrent units for temporal context modeling. Inspired by speech recognition advances, 
Conformer architectures \cite{convconformer} were adapted to SELD, with the ResNet Conformer \cite{resnetconformer} variant employing ResNet-18 encoders followed by Conformer layers to improve task-specific representation learning. Subsequently, 
CST-former\cite{cstformer}introduced attention mechanisms for cross-domain feature extraction and fusion, enabling the capture of complex spatio-temporal relationships. 

Further innovations in feature representation have also emerged. 
While traditional methods relied on basic spectrograms, 
more nuanced approaches now employ specialized time-frequency features. 
For instance, the DCASE 2021 SELDnet utilized log-Mel spectrograms alongside intensity vectors. 
G-SELD \cite{gseld} explored Gammatone filters, known for their noise resilience, 
as a replacement for Mel banks. Bark filters, which are perceptually aligned and widely used in speech processing \cite{bark02}, 
offer another promising alternative, though their application in SELD is still underexplored.

Our key contributions are: 1) establishing the 3D SELD system's performance on 1-second audio segments, 2) systematically comparing three widely used filter banks, and 3) introducing an SCConv-enhanced CST block that significantly improves model performance.
\section{Methods}
\subsection{Filter Banks Alternatives}
The Bark scale is an acoustic scale where perceptually equal intervals correspond to equal distances on the frequency axis. This scale has been applied in various tasks such as emotion recognition\cite{bark01} and echo/noise suppression \cite{bark02}, yielding competitive results. 
In contrast, the Gammatone filter \cite{gammatone} is modeled to mimic the frequency selectivity of the cochlea, providing a more detailed representation of auditory frequency responses through asymmetric impulse responses.
G-SELD\cite{gseld} introduced a gammatone filter bank to replace Mel filters, showing small but consistent improvements in all metrics over the baseline in ablation studies, even near optimal performance.

Building on this work, we conduct a systematic comparison of three filter banks: Bark, Mel, and Gammatone, within SELD frameworks. Our ablation experiments (detailed in Chapter \ref{experiments}) indicate that Gammatone filters yield the highest localization accuracy, even under short time-segment constraints. To the best of our knowledge, this is the first comprehensive evaluation of these three filter banks in 3D SELD tasks, providing new insights into optimal filter selection for such applications.
The data preprocessing process is shown in the figure \ref{fig:preprocessing}.
\subsection{SCConv CST Block}
We adapt the CST-Former\cite{cstformer} architecture for short-segment inputs and revisit the Channel-Spectro-Temporal (CST) block to evaluate the integration of the SCConv\cite{Scconv} module, examining its impact on CST block performance and efficiency. 
The structure of the proposed model is illustrated in the figure\ref{fig:model}.

\subsubsection{CST block}
The CST block integrates three distinct multi-head self-attentions (MHSA) to capture features across channel (C-MHSA), 
spectral (S-MHSA), and temporal (T-MHSA) dimensions \cite{cstprevious}. 
It employs two strategies for channel attention: Divided Channel Attention (DCA) and Unfolded Local Embedding (ULE). 
DCA treats input channels as batch dimensions, while ULE leverages temporal-frequency patches to enhance spatial and spectral representations, 
enabling unified spatio-spectral-temporal modeling.

Additionally, the CST block incorporates two convolutional components from Convolution Meets Transformer (CMT)\cite{cmt}: the Local Perception Unit (LPU) and Inverted Residual Feed-Forward Network (IRFFN), 
utilizing depth-wise convolutions to capture local features and refine representation.
\subsubsection{SCConv block}
The SCConv module comprises two units: the Spatial Reconstruction Unit (SRU) and the Channel Reconstruction Unit (CRU), designed to eliminate spatial and channel redundancies in CNN models. The SRU uses a Separate-and-Reconstruct strategy, leveraging Group Normalization\cite{group} to refine spatial features. The CRU adopts a Split-Transform-and-Fuse\cite{skt} approach to further reduce channel redundancies and maintain computational efficiency, making SCConv a plug-and-play module suitable for replacing standard convolutions in diverse architectures.

SCConv’s ability to refine spatio-temporal features makes it highly suitable for SELD tasks, where spatial and channel interactions are critical, enabling improved localization accuracy and robustness.
\subsubsection{SCConv CST block}
Our innovation involves replacing the original convolutional components—the LPU and IRFFN—with the SCConv module. This substitution mitigates spatial and channel redundancies, enhancing feature expressiveness while maintaining efficiency. Additionally, we incorporate a residual connection to facilitate feature propagation and improve stability\cite{residual}, preserving input and output dimensions to ensure compatibility with existing CST-based architectures, allowing for seamless integration into SELD frameworks.

\begin{figure}[t]  
    \centering  
    \includegraphics[width=0.5\textwidth]{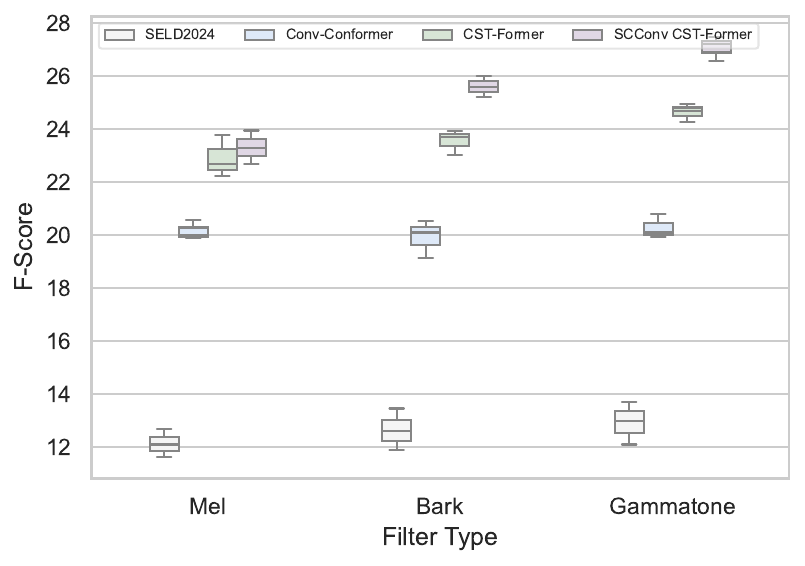}  
    \caption{Comparison of F-scores for four models (SELD2024, Conv-Conformer, CST-Former, and SCConv CST-Former) across three different filter types (Mel, Bark and Gammatone).} 
    \label{fig:fscore}  
\end{figure}
\subsection{Loss Function}
The proposed loss function follows the multi-ACCDOA format combined with the Auxiliary Duplicating Permutation Invariant Training (ADPIT) strategy, employing mean square error as the loss metric. 
This enables robust localization and detection by maintaining consistency across varying event orders, even in short time frames. 
The ADPIT formulation jointly optimizes localization and detection objectives, making it highly effective for polyphonic scenarios with overlapping sound sources. This loss function structure enhances SELD performance by accurately capturing spatio-temporal dynamics in complex acoustic environments.
\begin{equation} \mathcal{L} = \frac{1}{C T} \sum_{c}^{C} \sum_{t}^{T}  \min_{p \in P} 
l_{p,ct}
\label{}
\end{equation}

\begin{equation} l_{p,ct} = \frac{1}{N} \sum_{n}^{N}\text{MSE}
(\mathbf{P}_{p,nct}^{*}, \mathbf{\hat{P}}_{p,nct}^{*})
\label{}
\end{equation}
In these equations,  $N$  denotes the number of tracks, $C$  is the number of sound event classes, $T$ represents the number of time frames, and $ p \in P $ refers to a possible permutation. $\mathbf{P}_{p,nct}^{*}, \mathbf{\hat{P}}_{p,nct}^{*} \in \mathbb{R}^{3\times N\times C \times T}$ represent the ACCDOA target and prediction, respectively.

\begin{table*}[t]  
    \centering
    \caption{Model Parameters Comparison}
    \begin{tabular}{lccccc}  
    \toprule
    \textbf{Model}            & \textbf{Parameters (M)}  & $F_{20^\circ/1}\uparrow$ & $AE\downarrow$ & $RDE\downarrow$ & $\varepsilon_{SELD}\downarrow$ \\ \midrule
    SELD2024                  & 0.84                    & 12.6               & 24.4   & 0.38   & 0.45                  \\ 
    Conv-Conformer+MHSA                 & 16.50                  & 19.3             & 29.57   & 0.298   & 0.425                  \\ 
    \rowcolor{gray!20}
    Conv-Conformer            & 14.39                     & 20.7                  & 28.2   & \pmb{0.284}   & \pmb{0.413}                  \\ 
    CST Former                & 0.54                    & 24.7             & 23.0 & 0.699& 0.526               \\ 
    \rowcolor{gray!20}
    SCConv CST Former(ours)        & 0.57                    & \pmb{27.2}              & \pmb{22.4}  & 0.693   & 0.515            \\ \bottomrule
    \end{tabular}
    \label{tab:model_params}
\end{table*}
\section{Experimental Setup}
\subsection{Dataset and Preprocessing}
The dataset used is derived from the Synthetic SELD mixtures in DCASE2024 Task 3\cite{generatedataset}, using the First-Order Ambisonics (FOA) format. It consists of 1200 one-minute spatial recordings, split into 900 for training and 300 for validation and testing, totaling 72,000 samples. During preprocessing, intermediate features were extracted to minimize computational overhead. The label hop length was reduced from 100 ms to 50 ms for finer temporal resolution.

STFT was applied to each recording using a 512-point FFT, a 10-ms hop length, and a 20-ms window. Mel, Gammatone, and Bark filters were then used to generate non-linear spectrograms, which were concatenated with XYZ intensity vectors (excluding W), resulting in a feature map of dimensions (7, T, 128). This configuration balances representation detail with efficiency. Finally, we used PyTorch’s DataLoader to replace the original data generator, reducing training time by 50\% by shifting the bottleneck from data loading to model processing.

\subsection{Metrics}
We evaluated system performance using four primary metrics: thresholded F-score ($F_{20^\circ/1}$), Angular Error ($AE$), Relative Distance Error ($RDE$), 
and SELD Score ($\varepsilon_{SELD}$). 
A threshold of $T_{\text{DOA}} = 20^\circ$ and a relative distance threshold of $T_{\text{RD}} = 1$ are applied to define true positive matches, with results averaged over all event classes. The SELD Score ($\varepsilon_{SELD}$) is computed using four sub-metrics: Localization Error ($LE_{CD}$), Localization Recall ($LR_{CD}$), Error Rate ($ER_{20^\circ}$), and F-score ($F_{20^\circ}$), defined as:
\begin{equation}
\varepsilon_{SELD} = \frac{ER_{20^\circ}  + 2 -F_{20^\circ}  -LR_{CD} + \frac{LE_{CD}}{180^\circ
} }{4}.
    \label{}
\end{equation}

The optimal values for these metrics are: $F_{20^\circ/1} = 100\%$, $AE = 0^\circ$, $RDE = 0$, and $\varepsilon_{SELD} = 0$. We use the $F_{20^\circ/1}$ as the early stopping criterion with a patience of 50 epochs.

\subsection{Training Configuration and Baseline Methods}
A consistent experimental setup was used for all evaluations. The Adam optimizer with a learning rate of 0.0005 and a CosineAnnealingLR scheduler (maximum iterations, $T_{max} = 10,000$) was applied. PyTorch’s DataLoader handled training, while custom generators were used for validation and testing. All experiments were conducted on an NVIDIA V100 GPU.

The baseline models include the SELDnet of 2024 audio-only track, Conv-Conformer (with and without MHSA), and CST-Former adapted to match the short-segment input shape. Finally, SCConv was integrated with each of these methods to evaluate its impact.

\section{Results and Discussion}\label{experiments}

\subsection{Filterbank Evaluation}
To rigorously assess the impact of various filter banks on system performance, we conducted an evaluation using three commonly utilized filter types—Mel, Gammatone, and Bark. Each experiment maintained a consistent setup with a 1-second segment length and 128 frequency bins to ensure comparability across models. The results, as depicted in Figure \ref{fig:fscore}, demonstrate that the Gammatone filter consistently outperforms the Mel and Bark filters in terms of F-score, yielding average improvements of 11.3\% and 7.6\%, respectively. This superior performance is attributed to the Gammatone filter’s precise modeling of cochlear frequency selectivity, which allows for more accurate representation of auditory signals. By leveraging this characteristic, the Gammatone filter enhances feature extraction, particularly in complex auditory scenes with overlapping sources, ultimately improving polyphonic sound event localization and detection.

\subsection{Ablation Study: Impact of SCConv on CST Block}
To evaluate the integration of SCConv into the CST block, we compared the CST Former with the SCConv-enhanced version across all three filter types using a 1-second segment.
As shown in Table \ref{tab:filter_performance}, SCConv CST Former outperforms the original CST Former on nearly all localization metrics, demonstrating superior robustness. This is likely due to SCConv’s ability to mitigate spatial and channel redundancies, enhancing feature representation while maintaining computational efficiency.

\subsection{Comparative Benchmark Analysis}
We compared SELD2024, Conv-Conformer (with and without MHSA), CST Former, and SCConv CST Former using key performance metrics.
Despite a parameter count of only 0.5M, both CST Former and SCConv CST Former surpass the Conv-Conformer model, which has 14.4M parameters, in terms of threshold F-score.
Notably, SCConv CST Former consistently outperforms CST Former in all evaluated metrics, validating its effectiveness.
However, Conv-Conformer exhibits superior performance in Relative Distance Error (RDE) and SELD score, indicating its advantage in precise distance estimation and spatial localization, likely due to the enhanced model representation capabilities.

Future work could focus on simultaneously enhancing F-score and RDE through ensemble strategies or employing separate models for distinct predictions.

\begin{table}[t]  
    \centering
    \caption{SCConv CST former and CST former Performance Comparison of Different Filters}
    \renewcommand{\arraystretch}{0.9}  
    \small  
    \begin{tabular}{lp{1cm}p{0.6cm}p{0.6cm}p{0.6cm}p{0.6cm}}
    \toprule
    \multirow{2}{*}{Model}         & \multirow{2}{*}{Filter} & \multicolumn{4}{c}{Performance Metrics} \\ \cmidrule(lr){3-6} 
                                   &                        & $F_{20^\circ/1}$   & $AE$   & $RDE$   &  $\varepsilon_{SELD}$   \\ \midrule
    CST Former                     & \multirow{2}{*}{Mel}   &  22.7   &     24.2   &   0.701    & 0.534            \\ \cmidrule(lr){1-1} \cmidrule(lr){3-6} 
    SCConv CST Former              &                        &   23.3  &    23.3  &  0.700     &       0.530      \\ \midrule
    CST Former                     &\multirow{2}{*}{Bark} & 23.7  & 23.1   & 0.696    &      0.529    \\ \cmidrule(lr){1-1} \cmidrule(lr){3-6} 
    SCConv CST Former              &                        &  25.6   &   22.1   &   0.697    &     0.521        \\ \midrule
    CST Former                     & Gamma  &  24.7   &  23.0    & 0.699     &  0.526            \\ \cmidrule(lr){1-1} \cmidrule(lr){3-6} 
    SCConv CST Former              &   -tone  &  \pmb{27.2}   & \pmb{22.4}    &  \pmb{0.693}     &  \pmb{0.515}           \\ \bottomrule
    \end{tabular}
    \label{tab:filter_performance}  
\end{table}

\section{Conclusion}
This paper presents an evaluation of 3D SELD systems under short time-segment scenarios, 
addressing a critical gap in current SELD research and advancing the applicability of SELD systems by establishing a new baseline for 1-second segment analysis. 
We conducted a comparative analysis of three filter bank schemes in 3D SELD systems to inform filter selection for feature extraction. 
Additionally, by replacing convolutional components in the CST block with the SCConv module, 
we achieved consistent improvements across multiple 3D SELD metrics, highlighting SCConv’s potential in enhancing spatial and channel feature representations. 
and suggesting that SCConv can serve as a valuable component for future SELD system development. 

\clearpage
\bibliographystyle{unsrt}
\bibliography{references}

\vspace{12pt}
\color{red}

\end{document}